\documentclass[twocolumn,english]{revtex4-1}
\usepackage[T1]{fontenc}
\usepackage[latin9]{inputenc}
\usepackage{verbatim}
\usepackage{amsmath}
\usepackage{amssymb}
\usepackage{graphicx}
\usepackage{babel}
\begin{document}
\global\long\def\ket#1{\left|#1\right\rangle }

\global\long\def\bra#1{\left\langle #1\right|}

\global\long\def\braket#1#2{\left\langle #1\left|#2\right.\right\rangle }

\global\long\def\ketbra#1#2{\left|#1\right\rangle \left\langle #2\right|}

\global\long\def\braOket#1#2#3{\left\langle #1\left|#2\right|#3\right\rangle }

\global\long\def\mc#1{\mathcal{#1}}

\global\long\def\nrm#1{\left\Vert #1\right\Vert }

\title{Global passivity in microscopic thermodynamics }
\begin{abstract}
The main thread that links classical thermodynamics and the thermodynamics
of small quantum systems is the celebrated Clausius inequality form
of the second law. However, its application to small quantum systems
suffers from two cardinal problems: (i) The Clausius inequality does
not hold when the system and environment are initially correlated
- a commonly encountered scenario in microscopic setups. (ii) In some
other cases, the Clausius inequality does not provide any useful information
(e.g. in dephasing scenarios). We address these deficiencies by developing
the notion of global passivity and employing it as a tool for deriving
thermodynamic inequalities on observables. For initially uncorrelated
thermal environments the global passivity framework recovers the Clausius
inequality. More generally, global passivity provides an extension
of the Clausius inequality that holds even in the presences of strong
initial system-environment correlations. Crucially, the present framework
provides additional thermodynamic bounds on expectation values. To
illustrate the role of the additional bounds we use them to detect
unaccounted heat leaks and weak feedback operations (\textquotedblleft Maxwell's
demons\textquotedblright ) that the Clausius inequality cannot detect.
In addition, it is shown that global passivity can put practical upper
and lower bounds on the buildup of system-environment correlation
for dephasing interactions. Our findings are highly relevant for experiments
in various systems such as ion traps, superconducting circuits, atoms
in optical cavities and more.
\end{abstract}

\author{Raam Uzdin}

\author{Saar Rahav}

\affiliation{Schulich Faculty of Chemistry, Technion - Israel Institute of Technology,
Haifa, Israel}
\maketitle

\section{Introduction}

Recent years have seen a surge of interest in the thermodynamics of
small systems. Classical thermodynamics was developed for macroscopic
systems that are weakly coupled to large environments. Technological
advances now allow studying various processes in nanoscopic systems
and it is of great interest to understand such processes from a thermodynamic
point of view. 

Applying thermodynamics to processes in small systems requires going
beyond the assumptions and methodologies used in classical thermodynamics
for several reasons: (1) The dynamics of microscopic systems is quantum
and questions regarding the thermodynamic role of quantum coherence,
entanglement, and measurements become important; (2) The system-environment
coupling cannot generally be assumed to be weak. As a result the environment
is modified by the system, and strong recurrences may take place;
(3) A non-negligible degree of initial system-environment correlation
may be present, leading to effects such as heat flow from a cold subsystem
to a hot one \cite{Jennings2010ReverseFlow,SerraLutzReversingArrow2017};
(4) Small quantum systems are easily taken out of equilibrium and
therefore their dynamics cannot be efficiently described by small
number of quantities such as volume, average energy and so on. In
summary, it is of great interest to try to adapt thermodynamics to
deal with some of these deviations from the assumptions used in classical
macroscopic thermodynamics.

Several theoretical advances that extend thermodynamics were developed
in the last few decades. Stochastic thermodynamics \cite{Seifert2012StochasticReview,sekimotoStochEnergBook}
describes the fluctuations in thermodynamic characteristics of a process
such as heat and work. These were found to satisfy the celebrated
fluctuation theorems \cite{Jarzynski2011equalitiesReview,harris2007fluctuationReview},
a family of equalities that also hold far from equilibrium. Thermodynamic
resource theory \cite{GourRTreview,BrandaoPnasRT2ndLaw,horodecki2013fundamental,LostaglioRudolphCohConstraint}
studies the possible transformations a system can undergo by interacting
with a thermal bath. Both theories have their limitations. Stochastic
thermodynamics consider either systems decoupled from the environment
or systems coupled to macroscopic environments that remain in equilibrium
during the process. Resource theory is limited to a specific set of
operations called ``thermal operations''.

The interest in the thermodynamics of small system has led to several
experimental realizations of microscopic processes that take after
macroscopic thermodynamics. A heat engine with a single ion as a working
fluid \cite{rossnagelIonEngExp}, as well as a three-ion absorption
refrigerator \cite{Roulet_3_Ion_fridge2017} have been implemented.
Algorithmic cooling \cite{boykin2002algorithmic} has been demonstrated
in NMR \cite{AlgoCoolQuantGas2011Nature}. In addition, quantum features
of heat machines have also been recently observed in NV centers in
diamonds \cite{KlatzowPoemNVequiveExperiment}. An experiment demonstrating
the thermodynamic role of initial correlations was done in NMR \cite{SerraLutzReversingArrow2017}.
While these experiments verify the validity of the second law for
the smallest quantum systems, they also offer the possibility of testing
new thermodynamic predictions such as the ones suggested in this paper. 

In this work, we will mostly be interested in processes where a small
system of interest is coupled to other small systems that act as environments.
As a starting point the small environments are assumed to be initially
in Gibbs states $\rho_{0}^{(i)}=\exp(-\beta_{i}H_{i})/\text{tr}[\exp(-\beta_{i}H_{i})]$,
where $H_{i}$ is the Hamiltonian of environment $i$ and $\beta_{i}$
is its inverse temperatures. The system of interest can start in any
state. Due to the microscopic size of the small environments, the
interaction with the system will in general modify them, and they
can substantially deviate from their initial Gibbs state. Moreover,
they may develop a strong correlation to the system of interest. We
call such small and initially thermal environments 'microbaths' to
distinguish them from macroscopic baths encountered in classical thermodynamics. 

The dynamics of the whole setup containing the system of interest
and all the microbaths is described by a global unitary dynamics (quantum
evolution). More generally the dynamics can be described by a statistical
mixture of unitaries. Physically, a mixture of unitaries corresponds
to the scenario where there is some noise in the controls generating
the thermodynamic protocol. These global unitaries act on both the
system and the microbaths (leading to heat flows), and can also do
work. Our goal is to describe the thermodynamics of such processes
even when the elements are initially correlated to each other.

One of the characteristics of thermodynamics is the appearance of
inequalities. The second and third laws tell us that there are tasks
that cannot be done. In the current setup the most relevant form of
the second law is the Clausius inequality (CI). In the following,
we show that the concepts of passivity \cite{pusz78,lenard1978Gibbs,AllahverdyanErgotropy,MartiWorkCorr,WolgangSqueezedErgotropy}
can be extended and be used to derive additional inequalities. These
inequalities have several appealing features not exhibited by the
CI: (1) they hold for systems that have some initial quantum or classical
correlation to the environment; (2) They can set upper and lower bounds
on the system-environment correlation buildup; (3) They allow to derive
families of inequalities that can detect external tempering in the
form of heat leaks or feedback (e.g., a Maxwell demon) even when the
CI fails to detect the feedback.

The flexibility and general applicability of the global passivity
framework presented in this work come at a price. Global passivity
connects the initial state of the setup to the observable appearing
in the resulting inequalities. As a result, in some cases the predictions
obtained from these inequalities may involve non-intuitive observables.
Nevertheless, the examples we give in this paper demonstrate that
one can derive interesting new predictions in various important scenarios
that were so far outside the scope of the thermodynamic description.
We believe that the flexibility of this framework is useful, and will
lead to additional predictions on measurable quantities in nanoscopic
setups. 

\begin{figure}
\includegraphics[width=8.6cm]{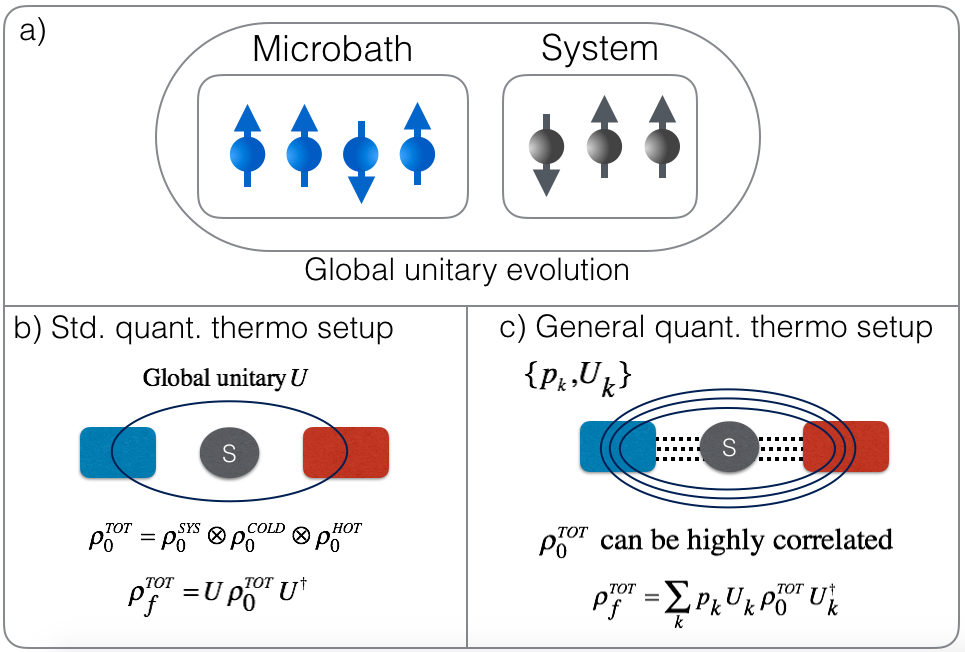}

\caption{(a) A typical scenario for our theory. A microbath with very small
heat capacity (e.g. several spins) is initially prepared in a thermal
state and then coupled to a system in a non-thermal state. In such
scenarios the dynamics is highly non-Markovian and in addition the
unitary transformation that generates the interactions may add or
remove energy from the system-environment setup. (b) In the standard
quantum thermodynamic setup all elements (microbaths and system) are
initially uncorrelated and then a global unitary (a thermodynamic
protocol) describes their interaction. (c) In this paper we consider
any initial conditions including strong entanglement between the system
and the microbaths. Moreover we allow for a mixture of unitaries which
include the possibilities of noise in the thermodynamic protocol.}
\end{figure}

In Sec. II we introduce the notion of global passivity and use it
to derive a version of the Clausius inequality that is valid in the
presence of initial system-environment correlation. Several important
examples are studied in detail. Section III uses the concept of global
passivity to obtain a new type of inequalities. We exemplify the use
of these new inequalities for detecting unaccounted heat leaks, for
detecting the presence of ``lazy'' Maxwell demons, and for studying
the buildup of system-environment correlation in a dephasing scenario.
In these examples, the results obtained from global passivity are
much more useful compared to the standard second law. We summarize
our findings in Sec. IV.

\section{Extending the Clausius inequality using global passivity}

In this the section, we introduce the notion of global passivity and
demonstrate how it can be used to derive various thermodynamic inequalities.
Before doing so, we present the celebrated Clausius inequality formulation
of the second law of thermodynamics. Under some restrictions, this
formulation holds in some microscopic setups. We discuss its structure
and its limitations when applied to microscopic systems. 

\subsection{The Clausius inequality in microscopic setups\label{sec: The CI}}

Historically the second law was developed for macroscopic systems
such as steam engines and large thermal reservoirs. However, it turns
out that under some conditions one of its formulations, the Clausius
inequality (CI) also holds for small quantum systems interacting with
each other and with external fields. Since this corresponds to the
setup considered in this paper, the CI will serve as a natural reference
for the new results described in this article.

Consider a setup in which a system of interest and several other small
systems are prepared in an initially uncorrelated state of the form
\begin{equation}
\rho_{0}^{tot}=\rho_{0}^{sys}\otimes e^{-\beta_{1}H_{1}}\otimes...e^{-\beta_{N}H_{N}}/Z_{0},\label{eq: basic thermo r0}
\end{equation}
where $Z_{0}=\text{tr}[e^{-\sum_{i=1}^{N}\beta_{i}H_{i}}]$ is a normalization
factor, $H_{i}$ are the Hamiltonians of the systems that act as microbaths,
where $\beta_{i}$ corresponds to the \textit{initial} inverse temperatures.
A thermodynamic process is realized via some time-dependent global
Hamiltonian that acts on all elements (not necessarily simultaneously).
As a result, the setup evolves unitarily in time. The initial preparation
distinguishes the system of interest, which can be in any initial
state, from the microbaths that are initially in a thermal state.
We note that this initial state is special due to the lack of correlation
between the different elements.

The process described above satisfies a quantum-microscopic version
of the CI \cite{Sagawa2012second,PeresBook,Esposito2011EPL2Law},
\begin{equation}
\Delta S^{sys}+\sum_{k}\beta_{k}q_{k}^{bath}\ge0,\label{eq: Basic CI}
\end{equation}
where $q_{k}=\Delta\left\langle H_{k}\right\rangle $ is the change
in the average energy of microbath $k$. $\Delta S^{sys}=S(\rho_{f}^{sys})-S(\rho_{0}^{sys})$
is the change in the von Neumann entropy of the system $S^{sys}(\rho^{sys})\doteq-\text{tr}[\rho^{sys}\ln\rho^{sys}]$
where $\rho^{sys}=\text{tr}_{baths}[\rho^{tot}]$. This microscopic
version goes beyond the assumptions of classical thermodynamics since
the microbaths can be arbitrary small in size and be driven far from
equilibrium during the process. However, it is restricted by the demand
that all the elements are initially uncorrelated and that the microbaths
start in thermal state. 

One of the goals of this paper is to generalize (\ref{eq: Basic CI})
to the case where the elements are initially correlated. To make sure
we find a plausible generalization we now take a closer look at the
structure of the CI. The first term deals with changes in a quantity
that is nonlinear in the density matrix $S(\rho)=-tr[\rho\ln\rho]$.
The non linearity appears both in the initial and the final state.
As such, it is not an observable that can be directly measured, but
rather a quantity that is calculated after $\rho$ has been evaluated
via tomography of the system. Nevertheless, its informational interpretation
makes it very useful. The second term $\sum_{k}\beta_{k}q_{k}^{bath}$
is a measurable quantity that describes changes in expectation values. 

With this combination of system information and observables, the CI
neatly expresses the energy-information relation that appears in fundamental
processes such as Landauer's erasure, Szilard engine, and reversible
state preparation \cite{Anders2015MeasurementWork,RU2017genCI}. When
extending the CI it is desirable to maintain this information-expectation
value structure. See \cite{RU2017genCI} for an extension of the second
law that preserves the information-observable structure, and \cite{horodecki2013fundamental,LostaglioRudolphCohConstraint}
for an extension that does not. 

Finally, we point out that using (\ref{eq: basic thermo r0}) and
$\rho_{0}^{env}\doteq\text{tr}_{sys}\rho_{0}^{tot}$, the term $\sum_{k}\beta_{k}q_{k}^{bath}$
can be written as $\text{tr}[(\rho_{f}^{env}-\rho_{0}^{env})(-\ln\rho_{0}^{env})]$.
Using the notation 
\begin{equation}
\mc B^{env}\doteq-\ln\rho_{0}^{env},\label{eq: B bath}
\end{equation}
 the standard CI (\ref{eq: Basic CI}) can be written in a form that
will be useful later
\begin{equation}
\Delta S^{sys}+\Delta\left\langle \mc B^{env}\right\rangle \ge0.\label{eq: CI Bbath}
\end{equation}

\subsection{Passivity and expectation values inequalities\label{subsec: passive-operators}}

Passivity \cite{pusz78,lenard1978Gibbs,AllahverdyanErgotropy,Marti2015EnergeticPassive,MartiWorkCorr,Wolgang2018passivityCI}
is defined as follows: a time-independent operator $\mathcal{A}$
and a density matrix $\rho$ are said to be passive w.r.t. each other
if (i) $\rho$ and $\mathcal{A}$ are diagonal in the same basis (same
eigenvectors). (ii) in a basis sorted in increasing order of eigenvalues
of $\mc A$, the eigenvalues of $\rho$ are decreasing. Since the
eigenvalues of $\rho$ correspond to probabilities it implies that
when measuring $\mathcal{A}$ in a system prepared in a passive state,
higher eigenvalues of $\mathcal{A}$ are less probable to be observed
than lower eigenvalues. This is illustrated in Fig. 2a.

Passive pairs $\{\mathcal{A},\rho_{\mathcal{A}}\}$ satisfy an important
inequality. Consider an initial passive state $\rho_{\mathcal{A}}$
w.r.t $\mathcal{A}$ that is mapped to a final state via a unitary
transformation

\begin{equation}
\rho_{f}=U\rho_{\mathcal{A}}U^{\dagger},\label{eq: rho f U}
\end{equation}
 or more generally by a mixture of unitaries

\begin{equation}
\rho_{f}=\sum_{k}p_{k}U_{k}\rho_{\mathcal{A}}U_{k}^{\dagger},\label{eq: rho f mixU}
\end{equation}
where $p_{k}$ denotes the probability of executing the unitary $U_{k}$.
The passivity of the initial state ensures that the expectation value
$\left\langle \mathcal{A}\right\rangle _{f}=\text{tr}(\rho_{f}\mathcal{A})$
satisfies 
\begin{align}
\left\langle \mathcal{A}\right\rangle _{f} & \ge\left\langle \mathcal{A}\right\rangle _{pass}=\text{tr}(\rho_{\mathcal{A}}\mathcal{A}).\label{eq: passivty ge}
\end{align}
Alternatively stated, the passive state gives the lowest expectation
value achievable by mixture of unitaries transformations. The validity
for mixture of unitaries follows from linearity and the validity for
any single unitary (\ref{eq: rho f U}). To see why (\ref{eq: passivty ge})
holds for any unitary we use the fact that when a unitary operates
on a diagonal matrix then it holds that $\vec{p}_{f}$, the diagonal
elements of the new density matrix, are related to the original diagonals
$\vec{p}_{0}$ via a mixture of permutations $\Pi_{k}$ that are executed
with some probability $\zeta_{k}$: $\vec{p}_{f}=\sum_{k}\zeta_{k}\Pi_{k}\vec{p}_{0}$.
When the initial distribution is passive w.r.t to $\mc A$, $\vec{p}_{0}=\vec{p}_{pass}$,
any permutation or a mixture of permutations (doubly stochastic map)
will increase the expectation value of $\mc A$ (Fig. 2b). Alternatively,
one can use the proof presented in Ref. \cite{AllahverdyanErgotropy}.
\begin{figure}
\includegraphics[width=8.6cm]{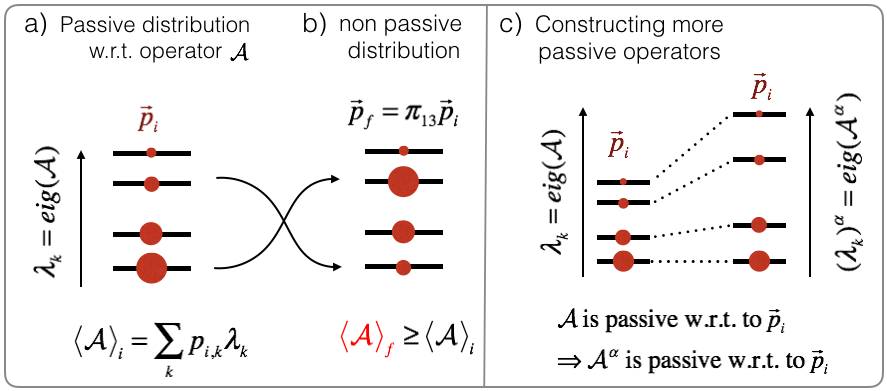}

\caption{Passivity of a general Hermitian operator $\protect\mc A$ with respect
to initial density matrix $\rho_{0}$. $\vec{p}_{i}$ denote the eigenvalues
of $\rho_{0}$. In passive distributions with respect to an observable
$\protect\mc A$, larger eigenvalues of $\protect\mc A$, (measurement
results $\lambda_{i}$) have lower probability to be observed (probability
is given by the size of the circles). Figure (b) illustrates that
starting from passive distribution, any permutation must increase
the expectation value $\left\langle \protect\mc A\right\rangle $.
This implies that $\left\langle \protect\mc A\right\rangle $ increases
under a unitary operation ($\Delta\left\langle \protect\mc A\right\rangle \ge0$).
(c) If $\protect\mc A$ is passive w.r.t. to $\vec{p}_{i}$, then,
$\protect\mc A^{\alpha}$ is passive w.r.t. $\vec{p}_{i}$ for any
$\alpha>0$.}
\end{figure}

Traditionally in thermodynamics, the passivity inequality is related
to the amount of work that can be extracted from a system using a
\textit{transient} unitary operation on the system. That is, the Hamiltonian
of the system is driven by some external fields for some time, but
in the end it returns to its original value (cyclic Hamiltonian).
This maximum extractable work is called the ergotropy of the system
\cite{AllahverdyanErgotropy}, and it is obtained in a unitary process
in which the system ends in a passive state with respect to the system
Hamiltonian. 

Among all the passive states with respect to the Hamiltonian, the
Gibbs state has a special status. The Gibbs state is the only state
that has the property of complete passivity w.r.t. the Hamiltonian
\cite{pusz78}. A collection of $N$ copies of the system in a Gibbs
state is also a passive state with respect to the total Hamiltonian.
Thus, no unitary that acts on the $N$ copies (including unitaries
that describe interactions between the copies) can be used to reduce
the total energy. This provides a link to the second law as it excludes
the work extraction from a single thermal reservoir. 

Motivated by work extraction, in previous studies of passivity \cite{pusz78,lenard1978Gibbs,AllahverdyanErgotropy,Marti2015EnergeticPassive,MartiWorkCorr,Wolgang2018passivityCI}
the operator $\mathcal{A}$ was the Hamiltonian of the system (or
of the environment \cite{Wolgang2018passivityCI}). The use of passivity
in this paper is different from the standard one in three different
ways: 1) Here, passivity is applied to the \textit{entire setup} and
not just to the system. We use the term 'global passivity' to distinguish
this scenario from the standard use of passivity that deal with work
extraction from a specific element in the setup. In particular, the
energy flows involved in global passivity include heat flows between
the elements. 2) We apply passivity to various operators not just
to Hamiltonians. 3) In traditional passivity the operator (the Hamiltonian)
is given and the passive states are the focus of interest. In contrast,
in this paper we ask ``\textit{given an initial density matrix $\rho_{0}$
what are the operators $\mathcal{A}$ that are passive with respect
to $\rho_{0}$?}''. The motivation is that for such an operator $\mathcal{A}$
it is guaranteed that
\begin{equation}
\Delta\left\langle \mathcal{A}\right\rangle =tr[(\rho_{f}-\rho_{0})\mathcal{A}]\ge0\label{eq: A r0 pass}
\end{equation}
for any $\rho_{f}$ generated from $\rho_{0}$ by a mixture of unitaries.
These three differences lead to a new connection between passivity
and the second law, and to new thermodynamic inequalities. 

Before we apply (\ref{eq: A r0 pass}) to the thermodynamic scenarios,
we describe two simple extensions of (\ref{eq: A r0 pass}) that are
used in this paper to obtain additional inequalities on expectation
values of observables. 

The first extension concerns operators which are \textit{not} passive
with respect to the initial state. Let $\rho_{f}$ be density matrix
obtained from $\rho_{0}$ by a mixture of unitaries. We are interested
in the change of the expectation value of an operator $\tilde{\mathcal{A}}$.
Even when the operator of interest $\tilde{\mathcal{A}}$ and the
initial state $\rho_{0}$ are \textit{not} passive with respect to
each other we can write
\begin{align}
\Delta\left\langle \tilde{\mathcal{A}}\right\rangle  & \doteq\left\langle \tilde{\mathcal{A}}\right\rangle _{f}-\left\langle \tilde{\mathcal{A}}\right\rangle _{0}\\
 & \ge\left\langle \tilde{\mathcal{A}}\right\rangle _{\min}-\left\langle \tilde{\mathcal{A}}\right\rangle _{0},
\end{align}
where $\left\langle \tilde{\mathcal{A}}\right\rangle _{\min}$ is
the minimal expectation value of $\tilde{\mathcal{A}}$ obtained by
transforming $\rho_{0}$ via a mixture of unitaries. From the definition
of passive states it holds that $\left\langle \tilde{\mathcal{A}}\right\rangle _{min}=tr[\rho_{\tilde{\mathcal{A}}\,pass}\tilde{\mathcal{A}}]\doteq\left\langle \tilde{\mathcal{A}}\right\rangle _{\tilde{\mathcal{A}}\,pass}$
where $\rho_{\tilde{\mathcal{A}}\,pass}$ is obtained from $\rho_{0}$
via a unitary rotation $V$ so that $\rho_{\tilde{\mathcal{A}}\,pass}$
is passive with respect to $\rho_{\tilde{\mathcal{A}}\,pass}$. Therefore,
we obtain
\begin{equation}
\Delta\left\langle \tilde{\mathcal{A}}\right\rangle \ge\left\langle \tilde{\mathcal{A}}\right\rangle _{\tilde{\mathcal{A}}\,pass}-\left\langle \tilde{\mathcal{A}}\right\rangle _{0}\label{eq: shifted reference}
\end{equation}
Importantly, the right hand side is independent of $\rho_{f}$. Thus,
it can be evaluated before running the experiment without knowing
the details of the evolution. As shown in Sec. \ref{subsec: dephasing}
Eq. (\ref{eq: shifted reference}) can be quite useful.

The second extension of passivity we exploit in this paper concerns
the case where a passive operator with respect to the initial density
matrix is given, and we want to find additional passive operators
in order to set additional constraints on additional observable quantities.
Let $\mathcal{A}$ be a passive operator w.r.t. to $\rho_{0}$, and
$g$ be an analytic and monotonically increasing function in the interval
between the smallest and largest eigenvalues of $\mathcal{A}$. It
holds that the operator $g(\mathcal{A})$ is also passive w.r.t. $\rho_{0}$
and therefore
\begin{equation}
\Delta\left\langle g(\mathcal{A})\right\rangle \ge0\label{eq: gA r0 pass}
\end{equation}
for any mixture of unitaries. As will be demonstrated later these
new inequalities are not just restatement of (\ref{eq: A r0 pass})
and they have different physical content. The proof of (\ref{eq: gA r0 pass})
immediately follows from the fact that $\mathcal{A}$ and $g(\mathcal{A})$
have the same eigenvectors and also the same eigenvalue ordering since
monotonically increasing functions just ``stretches'' the spectrum
(Fig. 2c) but do not change the order of the eigenvalues. As a result,
any density matrix $\rho_{0}$ that is passive with respect to $\mathcal{A}$
is also automatically passive with respect to $g(\mathcal{A})$.

\subsection{Global passivity and its relation to Clausius inequality}

To introduce the notion of global passivity and its relation to the
second law we first examine a simplified scenario. In the next subsection
we treat the general case. Consider a setup that includes only initially
uncorrelated microbaths. That is, there is no system that is initially
in a non thermal state. The initial state of the whole setup is therefore
given by 
\begin{equation}
\rho_{0}^{tot}=e^{-\beta_{1}H_{1}}\otimes...e^{-\beta_{N}H_{N}}/Z_{0},\label{eq: only microbs}
\end{equation}
where $Z_{0}=tr[e^{-\beta_{1}H_{1}}\otimes...e^{-\beta_{N}H_{N}}]$
is a normalization factor. This situation can describe an absorption
refrigerator such as the one implemented in \cite{Roulet_3_Ion_fridge2017}.
In this case the CI reduces to
\begin{equation}
\sum_{k}\beta_{k}q_{k}^{bath}\ge0.\label{eq: CI no sys}
\end{equation}
To apply passivity for this setup we look for operators that are passive
with respect to $\rho_{0}^{tot}$. A simple and systematic way to
achieve this is to use $\rho_{0}^{tot}$ for the construction of the
passive operators. We start with the elementary choice 
\begin{equation}
\mc B^{tot}\doteq-\ln\rho_{0}^{tot}.\label{eq: Btot def}
\end{equation}
We emphasize that (\ref{eq: Btot def}) defines a \textit{time-independent}
operator based on the initial state. In particular, the expectation
value of this operator at time t, reads $\left\langle \mc B^{tot}\right\rangle =\text{tr}[\rho^{tot}(t)(-\ln\rho_{0}^{tot})]$.
It is simple to verify that $\mc B^{tot}$ is $\rho_{0}^{tot}$ passive.
By inverting (\ref{eq: Btot def}) we get $\rho_{0}^{tot}=e^{-\mc B^{tot}}$
that immediately implies that larger eigenvalues of $\mc B$ are associated
with lower probabilities. 

Next, the setup evolves by a global unitary or mixture of unitaries
\begin{equation}
\rho_{f}^{tot}=\sum_{k}p_{k}U_{k}\rho_{0}^{tot}U_{k}^{\dagger}.\label{eq: rho_f tot mix}
\end{equation}
From passivity of $\hat{B}^{tot}$ w.r.t $\rho_{0}^{tot}$ we get
the inequality
\begin{equation}
\Delta\left\langle \mc B^{tot}\right\rangle \ge0.\label{eq: basic PI}
\end{equation}
For the specific initial state (\ref{eq: only microbs}) $\mc B^{tot}=\beta_{k}H_{k}-(\ln Z_{0})I$
where $I$ is the identity operator. As a result (\ref{eq: basic PI})
becomes $\sum\beta_{k}q_{k}\ge0$, and we retrieve the result predicted
from the CI for this setup. 

As mentioned earlier, we call this approach global passivity since
we apply passivity for the whole setup and not just to the system
of interest. In standard passivity, the operator is fixed (The Hamiltonian
of the system) and the quantity of interest is the passive state.
Here the state is fixed/given by the initial preparation of the setup,
and we are interested in operators that are passive with respect to
it. In that sense, the present approach is complementary to the one
usually used in the literature on passivity \cite{pusz78,lenard1978Gibbs,AllahverdyanErgotropy,MartiWorkCorr,Marti2015EnergeticPassive,Wolgang2018passivityCI}.

The global passivity inequality (\ref{eq: basic PI}) is valid also
for a setup that includes a system with an arbitrary initial state
(\ref{eq: basic thermo r0}). The resulting inequality is less tight
than the CI. On the other hand, it involves quantities that are easier
to evaluate since all quantities are linear in the final density matrix.
Hence there is no need to perform a full system tomography in order
to evaluate $S^{sys}(\rho_{f})$ which appears in the CI. Details
can be found in Appendix I.

Next, to obtain the complete CI form (\ref{eq: Basic CI}) that include
the system entropy term, we introduce the passivity-divergence relation.
Quite remarkably we obtain that the passivity-divergence relation
not only reproduces the full CI but also yields a version of the CI
that is valid in the presence of initial system-environment correlation. 

\subsection{The passivity-divergence relation and the correlation compatible
Clausius inequality}

In thermal machines and in various thermal processes (e.g. thermal
state preparation \cite{Anders2015MeasurementWork}) there is often
a system of interest that does not start in a thermal state. In such
cases, the inequality (\ref{eq: CI no sys}) is not applicable and
(\ref{eq: Basic CI}) has to be used. We now show to use a more powerful
version of global passivity to obtain an inequality that distinguishes
between a system and its environment. We emphasize that this is not
just a re-derivation of the CI since the inequality we obtain can
handle initial correlation between the system and the environment
(in contrast to the CI). 

Let us consider a setup initially prepared in a state $\rho_{0}^{tot}$.
In contrast to the assumption of lack of correlation between the system
of the environment (\ref{eq: basic thermo r0}) we now allow for a
general density matrix $\rho_{0}^{tot}$ that may contain classical
and quantum correlation between the various elements of the setup.
The initial reduced state of the system is obtained by tracing out
the environment $\rho_{0}^{sys}=\text{tr}_{env}\rho_{0}^{tot}$. Similarly
the initial state of the environment is $\rho_{0}^{env}=\text{tr}_{sys}\rho_{0}^{tot}$.

Our starting point is the following identity
\begin{align}
\text{tr}[(\rho_{2}-\rho_{1})(-\ln\rho_{1})] & \equiv S(\rho_{2})-S(\rho_{1})+D(\rho_{2}|\rho_{1}),\label{eq: rel ent ident}
\end{align}
where 
\begin{equation}
D(\rho_{2}|\rho_{1})\doteq\text{tr}[\rho_{2}(\ln\rho_{2}-\ln\rho_{1})]\ge0,
\end{equation}
is the quantum relative entropy. To connect it to the notion of global
passivity we set $\rho_{2}=\rho_{f}^{tot},\rho_{1}=\rho_{0}^{tot}\equiv\exp(-\mc B^{tot})$
in (\ref{eq: rel ent ident}) and obtain
\begin{equation}
\Delta\left\langle \mc B^{tot}\right\rangle \equiv\Delta S^{tot}+D(\rho_{f}^{tot}|\rho_{0}^{tot}).\label{eq: Btot ident}
\end{equation}
We now focus on the first term in the right hand side that describes
the change of the total von Neumann entropy of the setup. If the evolution
of the setup is given by an exact unitary transformation $\rho_{f}^{tot}=U\rho_{0}^{tot}U$
then $\Delta S^{tot}=S(\rho_{f}^{tot})-S(\rho_{0}^{tot})=0$\footnote{In small quantum systems the coarse-graining is obtained by the partial
trace operation that leads to a reduced (local) description. This
reduced description ignores the presence of the correlation between
the elements so the sum of entropies is not fixed in time. However,
the von Neumann of the whole setup $S(\rho^{tot})$, contains the
correlations, and since it is unitarily invariant it holds that $S(\rho_{f}^{tot})=S(U\rho_{0}^{tot}U^{\dagger})=S(\rho_{0}^{tot})$.}. When a mixture of unitaries (\ref{eq: rho_f tot mix}) is applied
to the setup, it holds that $\Delta S^{tot}\ge0$. This follows from
the fact that the von Neumann entropy is Schur concave \cite{Marshall1979MajorizationBook},
and $\rho_{0}^{tot}$ majorizes $\rho_{f}^{tot}$ (the majorization
follows immediately from (\ref{eq: rho_f tot mix})). Physically,
a mixture of unitaries describes the dynamics in the presence of some
noise/randomness in the protocol.

Using $\Delta S^{tot}\ge0$ in (\ref{eq: Btot ident}) we obtain the
passivity-divergence relation 
\begin{equation}
\Delta\left\langle \mc B^{tot}\right\rangle \ge D(\rho_{f}^{tot}|\rho_{0}^{tot}).\label{eq: SPI}
\end{equation}
Since $D(\rho_{f}^{tot}|\rho_{0}^{tot})\ge0$, the passivity divergence
relation (\ref{eq: SPI}) immediately implies the global passivity
$\Delta\left\langle \mc B^{tot}\right\rangle \ge0$. This is alternative
proof of the global passivity inequality. 

Relation (\ref{eq: SPI}) is expressed using quantities that involve
the whole setup. To derive an inequality that distinguish between
system and environment and has a clear connection to the CI we use
a property of the quantum relative entropy: the quantum relative entropy
decreases when subsystems are traced out \cite{nielsen2002QuantInfo}.
In particular, it holds that
\begin{equation}
D(\rho_{f}^{tot}|\rho_{0}^{tot})\ge D(\rho_{f}^{sys}|\rho_{0}^{sys})\label{eq: reduced rel ent}
\end{equation}
for \textit{any} $\rho_{f}^{tot},\rho_{0}^{tot}$ (even if the system
is correlated to the rest of the setup). As a result we get
\begin{equation}
\Delta\left\langle \mc B^{tot}\right\rangle \ge D(\rho_{f}^{sys}|\rho_{0}^{sys}).\label{eq: Btot Dsys}
\end{equation}
Applying (\ref{eq: rel ent ident}) to the system density matrix and
using it in (\ref{eq: Btot Dsys}), we obtain the Correlation (compatible)
Clausius Inequality (CCI)
\begin{equation}
\Delta S^{sys}+\Delta[\left\langle \mc B^{tot}\right\rangle -\left\langle \mc B^{sys}\right\rangle ]\ge0,\label{eq: CCI}
\end{equation}
where $\mc B^{sys}$ is defined as 
\[
\mc B^{sys}\doteq-\ln\rho_{0}^{sys}.
\]

The CCI immediately reduces to the CI when the system and the environment
are initially uncorrelated since in this case $\left\langle \mc B^{tot}\right\rangle -\left\langle \mc B^{sys}\right\rangle =\left\langle \mc B^{env}\right\rangle $.
The CCI (\ref{eq: CCI}) is one of the main results in this paper.
We use the terminology correlation compatible Clausius inequality
for two reasons. Firstly, it reduces to the CI when $\rho_{0}^{tot}=\rho_{0}^{sys}\otimes\rho_{0}^{env}$.
Secondly, the CCI has the same structure of the CI: it relates changes
in the von Neumann entropy of the system of interest and changes in
expectation values. The difference is that on top of the changes in
the environment expectation values, there is also a change in expectation
value related to the initial system-environment correlation.

The CCI can also be recast in a slightly different form
\begin{align}
\Delta S^{sys}+\Delta\left\langle \mc B^{env}\right\rangle +\Delta\left\langle \mc B^{corr}\right\rangle  & \ge0\label{eq: ECI corr}
\end{align}
where
\begin{align}
\mc B^{corr} & \doteq\mc B^{tot}-\mc B^{sys}\otimes I^{env}-I^{sys}\otimes\mc B^{env}\label{eq: Bcorr deff}
\end{align}
where $I$ is the identity operator, and $\mc B^{corr}$ is an Hermitian
operator that identically vanishes when the system and environment
are initially uncorrelated. As the other $\mc B$ operators, $\mc B^{corr}$
is a time-independent operator that is determined by $\rho_{0}^{tot}$. 

There is a 'price' for incorporating the initial sys-env correlation
into the second law in the form of expectation values. For the same
initial reduced state of the environment, the form of $\mc B^{corr}$
will depend on the initial correlations in $\rho_{0}^{tot}$. That
is, for different correlations, different expectation values appear
in the CCI. To put this 'price' in perspective, it is important to
note that the relation between the initial state of the setup and
the expectation values that appears in the CCI is already present
in the standard CI. The initial preparation of the environment in
Gibbs states leads to the $\sum_{k}\beta_{k}q_{k}$ term in the CI.
In the CCI we get that not only the initial state of the environment
determines the expectation values but also the correlation to the
system. 

The role of the new correlation operator can be experimentally studied
in various experimental setups such as ion traps, superconducting
qubits and more. One of the simplest physical scenarios where this
term strongly manifests is heat flow between two initially correlated
spins. This has recently experimentally demonstrated with NMR spins
with quantum correlation \cite{SerraLutzReversingArrow2017}. The
quantum correlation of the initial state manifests in non-zero geometric
quantum discord. The initially colder spin ('c'), will constitute
the system and the initially hotter ('h') will constitute the environment.
Due to the initial correlation heat can flow from the cold spin to
the hot spin and make the cold spin colder and the hot spin hotter. 

To test the CCI we use the experimental parameters of \cite{SerraLutzReversingArrow2017}
and plot (Fig. 3b) the CI accumulated entropy production: $\Delta S^{c}+\beta q_{h}$
and the LHS of the CCI $\Delta S^{c}+\beta q_{h}+\Delta\left\langle \mc B^{corr}\right\rangle $.
When there is no initial correlation both expression are identical
as shown in Fig. 3b. In the presence of correlation, the standard
entropy production $\Delta S^{c}+\beta q_{h}$ becomes negative as
the CI no longer holds for this scenario. In contrast, the CCI expression
$\Delta S^{c}+\beta q_{h}+\Delta\left\langle \mc B^{corr}\right\rangle $
remains positive at all times. 

\begin{figure}
\includegraphics[width=8.6cm]{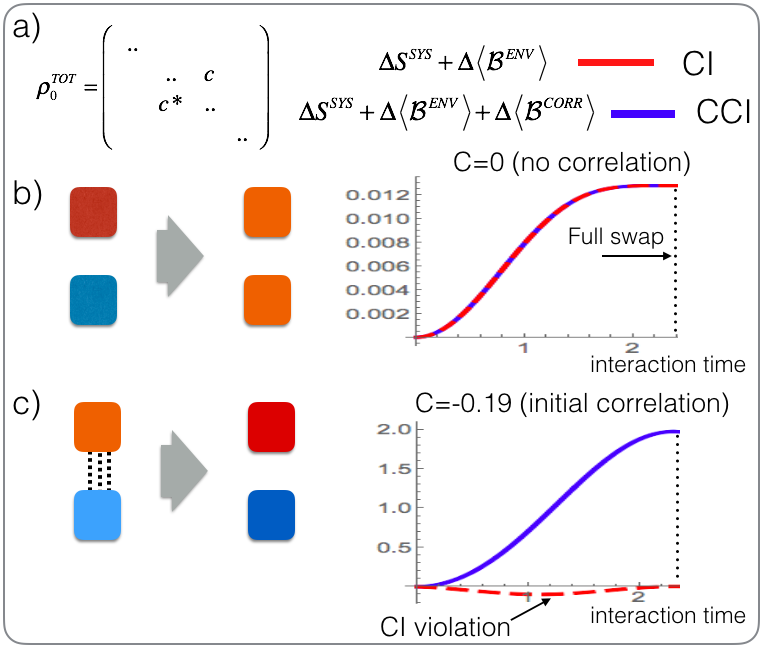}

\caption{\label{fig: ECI vs CI}The passivity-divergence relation offers an
extension of the Clausius inequality (second law) that can handle
initial correlation between environments or systems. (a) Hot and cold
qubits are prepared with initial correlation in the form of coherence
(superposition) between states $\protect\ket{01}$ and $\protect\ket{10}$
as shown by the form of the initial density. (b) In the absence of
initial correlation ($C=0$), an energy conserving interaction makes
the hot spin colder and the cold spin hotter. (c) In the presence
of initial correlation $C=-0.19$, the same interaction makes the
cold spin colder and the hot one hotter. All parameters are taken
from a recent NMR experiment \cite{SerraLutzReversingArrow2017}.
The top graph show that the CI ``entropy production'' (red) and
the left hand side (l.h.s) of the CCI (blue) have the same values
for this case and both are positive at all time. However in presence
of initial correlation (bottom graph), the l.h.s. of (\ref{eq: Basic CI})
attains negative values, and the CI fails. In contrast, the CCI remains
valid (positive l.h.s.) at all times. }
\end{figure}

Finally we emphasize that the essence of the CCI is the inequality
$D(\rho_{f}^{tot}|\rho_{0}^{tot})\ge D(\rho_{f}^{sys}|\rho_{0}^{sys})$
and not $\Delta S^{tot}\ge0$. $\Delta S^{tot}>0$ is simply another
layer of randomness that is externally added by the noise in the controls.
By using low noise control devices, an experimentalist can ensure
that the evolution is carried out by a single unitary to very high
accuracy. We also point out that the \textit{CCI is a non-perturbative
result} and is \textit{valid for} \textit{arbitrarily strong initial
correlations} and/or interactions between the elements.

In the next subsection, we study an important thermodynamic scenario
and work out explicitly the various terms in the CCI. As side comment,
we point out that the derivation above for the case $\rho_{0}^{tot}=\rho_{0}^{sys}\otimes\rho_{0}^{env}$
constitutes an alternative to the previous derivations of the standard
CI second law for microscopic quantum systems \cite{Sagawa2012second,PeresBook,Esposito2011EPL2Law,JarzynskiMicroscopicClausius}
(for the classical mechanics version see \cite{JarzynskiMicroscopicClausius}).

\subsection{CCI for a coupled thermal state\label{subsec:A-joint-system-bath}}

Next, we study an important case where the CCI can be written more
explicitly. Key quantities such as the interaction energy and the
potential of mean force will emerge naturally in the derivation. Moreover,
in Appendix II the structure obtained in this section is shown to
be valid in much more general circumstances. 

Let us consider a system that is permanently coupled to one of the
small environments available in the setup by some interaction Hamiltonian
$H_{I,0}$. When preparing this small environment in a thermal state
with inverse temperature $\beta_{h}$ (e.g. by weakly coupling it
to a larger thermal bath that is later entirely decoupled from the
setup), the system is also affected so the resulting sys-env state
is $\rho_{0}^{hs}=\frac{1}{Z_{hs}}e^{-\beta_{h}(H_{h}+H_{s}+H_{I,0})}$.
$H_{s}$ and $H_{h}$ are the bare Hamiltonians of the system and
the small environment respectively. Note that the \textit{reduced
state} of the system or the environment is generally not Boltzmann
distributed. 

For generality let us include another initially uncorrelated microbath
at inverse temperature $\beta_{c}$. Thus, the initial density matrix
is
\begin{equation}
\rho_{0}^{tot}=\frac{1}{Z_{hs}Z_{c}}e^{-\beta_{h}(H_{h}+H_{s}+H_{I,0})}e^{-\beta_{c}H_{c}}.\label{eq: couple thermal r0}
\end{equation}
For this setup the CCI (\ref{eq: CCI}) yields the following inequality

\begin{align}
\Delta S^{sys}+\beta_{c}q_{c}+\beta_{h}q_{h}+ & \beta_{h}[\Delta\left\langle H_{I,0}\right\rangle +\Delta\left\langle H_{s}-H_{s}^{\text{eff}}\right\rangle ]\nonumber \\
 & \ge0.\label{eq: ECI pre_fixed coupling}
\end{align}
As before $q_{c},q_{h}$ are the change in the average ``bare''
energies of the microbaths $\left\langle H_{c}\right\rangle ,\left\langle H_{h}\right\rangle $
and
\begin{equation}
H_{s}^{\text{eff}}\doteq-\frac{1}{\beta_{h}}\ln\rho_{0}^{sys}=-\frac{1}{\beta_{h}}\ln\text{tr}{}_{h}e^{-\beta_{h}(H_{s}+H_{I,0}+H_{h})}
\end{equation}
The quantity $H_{s}^{\text{eff}}-H_{s}$ is known as the potential
of means force \cite{kirkwood1935PMF} or as the solvation potential
\cite{Jarzynski2017_PRX_strong_coupling}. The first three terms in
(\ref{eq: ECI pre_fixed coupling}) are the ``bare'' terms that
appear in the Clausius inequality. The fourth term $\Delta\left\langle H_{I,0}\right\rangle $
originates from the CCI and it expresses the change in the interaction
energy during the process. The last term in (\ref{eq: ECI pre_fixed coupling})
$\Delta(H_{s}-H_{s}^{\text{eff}})$ is the change in the ``dressing
energy'' of the system. Due to the non-negligible interaction with
the environment the system is not initially in the thermal state of
the bare system Hamiltonian. 

While this paper was reviewed it was brought to our attention that
the CCI for the coupled thermal state has some similarity to a recent
classical formulation of the second law \cite{Jarzynski2017_PRX_strong_coupling}
for a very similar scenario. Despite the similarity there are quite
a few differences between our result (\ref{eq: ECI pre_fixed coupling}),
and \cite{Jarzynski2017_PRX_strong_coupling}. We point out the two
most important ones. First, our result is valid also in the presence
of coherences (in the energy basis) and quantum correlations that
arise from the initial system-environment coupling. Second, CI-like
inequalities that are derived from fluctuation theorems \cite{SeifertPRL2016_strong_coupling,JanetPRE2017_strong_coupling}
involve the \textit{equilibrium} entropy and \textit{equilibrium}
free energy. In contrast, Our result involve the von Neumann entropy
of the non-equilibrium state, and its corresponding \textit{non-equilibrium}
free energy \cite{CrooksThemoPredNonEf,Esposito2011EPL2Law}. 

In Refs. \cite{Winters2017correlation2ndLaw,reeb2014improved} appealing
treatments of initial correlations were presented in terms of changes
in mutual information and conditional entropy. Such quantities cannot
be expressed as expectation values (observables), and their evaluation
requires costly, or even impractical, system-environment tomography.
In contrast, the CCI the initial correlation manifests only in terms
measurable expectation values. Furthermore, Ref. \cite{Winters2017correlation2ndLaw}
assumes large environment, the availability of an external ancilla,
and the feasibility of global operations on many copies of the setup.
None of these assumptions are needed for the CCI. Finally, \cite{Winters2017correlation2ndLaw}
requires locally thermal states and therefore cannot handle the important
initial states such as the coupled thermal studied above (\ref{eq: couple thermal r0}).
In Ref. \cite{Jennings2015FTcorr} an interesting exchange fluctuation
theorem is suggested. However, this approach is valid only for systems
that are initially in local equilibrium. Moreover, it has no entropy
term as in the CI and the CCI. Thus, while \cite{reeb2014improved,Jennings2015FTcorr,Winters2017correlation2ndLaw}
constitute important contributions to the field, the findings in these
studies do not overlap with the results presented in this section.
In particular, in Sec. \ref{sec: alpha PSL} we derive additional
inequalities that were not obtained in Refs. \cite{Winters2017correlation2ndLaw,reeb2014improved}.

Equation (\ref{eq: ECI pre_fixed coupling}) was derived a setup prepared
in a coupled thermal state (\ref{eq: couple thermal r0}). Its form
seems to be rather different from the more general CCI (\ref{eq: CCI}).
However, in appendix II we show that the same form also exists for
a general initial state. In Appendix III we study in more detail two
physically interesting system-environment interactions. The first
is an energy conserving swap-like interaction, and the second is a
simple dephasing interaction. 

\section{additional passivity inequalities and their application\label{sec: alpha PSL}}

\subsection{Using passivity to generate new thermodynamic inequalities}

In Sec. \ref{subsec: passive-operators} it was shown that if an operator
$\mc A$ is passive w.r.t $\rho_{0}$, then $g(\mc A)$ is also a
passive operator w.r.t. $\rho_{0}$ where $g(x)$ is a monotonically
increasing function in the domain covering the eigenvalues of $\mc A$.
As shown before the operator $\mc B_{tot}=-\ln\rho_{0}$ is passive
w.r.t. to $\rho_{0}^{tot}$ for any thermodynamic protocol. Its eigenvalues
are non-negative. The function $g(x)=x^{\alpha}$ is an increasing
function for $x\ge0$ and $\alpha>0$. As a result, we get a new family
of global passivity inequalities
\begin{equation}
\Delta\left\langle \mc B^{\alpha}\right\rangle \ge0\label{eq: alpha GPI}
\end{equation}
for any $\alpha>0$ and for any thermodynamic protocol (\ref{eq: rho f mixU}).
For brevity in Sec. \ref{sec: alpha PSL} 'tot' is omitted and $\mc B$
will always refer to the whole setup. The $\alpha=1$ case was shown
earlier to coincide with the CI in the case of initially uncoupled
microbaths that interact with each other. Below we demonstrate using
three examples that the inequalities (\ref{eq: alpha GPI}) have different
physical content for different values of $\alpha$. This means that
inequalities with $\alpha\neq1$ carry information that is absent
in the CI. We also note that different $\alpha$ inequalities obey
a hierarchical structure as shown in Appendix IV.

\subsection{\label{subsec:Heat-leaks-detection}Heat leaks detection}

Let us assume that we are given a quantum chip (e.g. superconducting
circuit with several qubits). The unitary operation that the chip
is supposed to carry out is unknown to us, but we do have information
on the number of qubits, qudits and so on, that constitute the input
and output of the device. The thermodynamic challenge we face is to
verify if this chip is isolated from the surroundings (for all practical
purpose), or perhaps it interacts with some unaccounted for (hidden)
heat bath, e.g., via spontaneous emission or thermalization to the
substrate temperature. Note that the unitary implemented by the chip
is in general not energy-conserving (involves work). Hence, energy
conservation cannot be used to detect the heat leak to the hidden
environment.

One possible approach is to perform state tomography of the final
state of the setup and check if the eigenvalues of the density matrix
have been modified by an external agent. However, tomography of several
qubits is very hard to perform and impractical for dozens of qubits.
This method of detecting heat leaks is therefore very costly.

Another possibility is to use the second law. We prepare the input
state in a product of thermal states $\rho_{0}^{tot}=e^{-\sum\beta_{i}H_{i}}/Z$.
According to the CI (\ref{eq: Basic CI}) the final density matrix
satisfies $\sum_{i}\beta_{i}tr[\rho_{f}^{tot}H_{i}]\ge\sum_{i}\beta_{i}tr[\rho_{0}^{tot}H_{i}]$.
Indeed, if this inequality does not hold, one can deduce that there
is a heat leak in progress. Crucially, if the CI is not violated,
no conclusion can be made, as it possible that the heat leak is too
small to be detected by the CI. 

The global passivity inequalities (\ref{eq: alpha GPI}) suggest a
third possibility. A heat leak from the setup (chip) to some unaccounted
environment cannot be described as a mixture of unitary on the setup
only. Thus, in the presence of a heat leak from the chip, inequalities
(\ref{eq: alpha GPI}) may be violated and indicate the presence of
a heat leak. 

\begin{figure}
\includegraphics[width=8.6cm]{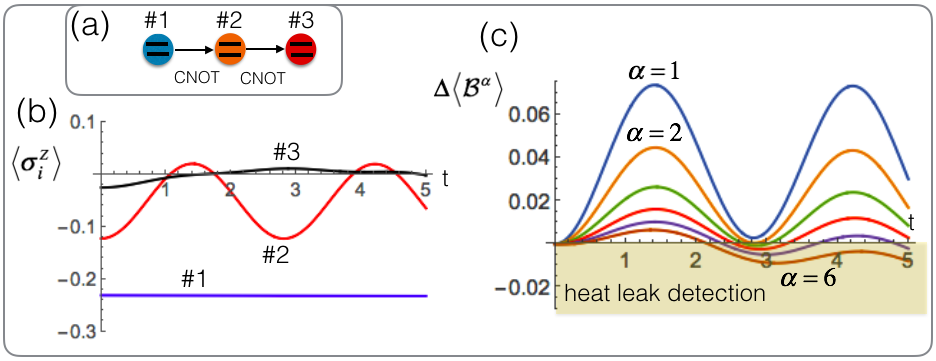}

\caption{\label{fig: heat leak}(a) Heat leak detection in a three-qubit circuit
with two CNOT interactions. The qubits also undergo decay to the ground
state at a rate $\gamma=1/1000$. This weak decay is not visible in
the polarization dynamics shown in (b). For example, the polarization
of qubit 1 (blue) looks perfectly flat. Figure (c) shows a validity
check of the inequalities (\ref{eq: alpha GPI}). Since the $\alpha=1$
curve that corresponds to the CI is always positive for the depicted
time, it cannot be used to detect the heat leak. On the other hand
the $\alpha=5$ and $\alpha=6$ global passivity inequalities are
violated in less the 3/1000 of the decay time (which is $1/\gamma=1000)$.}

\end{figure}

As an example, we study a three-qubit circuit with CNOT interaction
Hamiltonian between qubits 1 and 2, and qubits 2 and 3 (Fig. \ref{fig: heat leak}a).
The CNOT interaction strength is $\epsilon=1$ so the CNOT operation
is implemented in a time $\hbar/\epsilon$. The two CNOT interactions
operate simultaneously. In addition, there is a heat leak to a zero
temperature bath (e.g. spontaneous emission) at a rate $\gamma=\epsilon/1000$.
That is, a thousand CNOT cycles pass by before the decay becomes dominant.
Nevertheless, our goal is to detect the heat leak as soon as possible
since the chip typically operates only for a very short time with
respect to $1/\gamma$. 

The initial density matrix is $\rho_{0}^{tot}=e^{-\sum\beta_{i}\sigma_{z}^{(i)}}/Z$
with $\beta_{1}=1,\beta_{2}=0.5,\beta_{3}=0.1$. Fig. \ref{fig: heat leak}b
shows that for the depicted time the effect of the decay on the polarization
of the qubits is not visible. In Fig. \ref{fig: heat leak}c we plot
the values of $\Delta\left\langle B^{\alpha}\right\rangle $ as function
of time for various $\alpha$ values. When the curves take negative
values the heat leak is detected. The plot shows that larger $\alpha$
values can detect the heat leak much sooner compared to the $\alpha=1$
(which in this case equivalent to the CI).
It is important to emphasize that to evaluate $\Delta\left\langle \mc B^{\alpha}\right\rangle $
only local energy measurements $\{E_{1},E_{2},E_{3}\}$ are needed
in each realization (in each run of the experiment). $\left\langle \mc B^{\alpha}\right\rangle $
is obtained by the mean value of $(\beta_{1}E_{1}+\beta_{2}E_{2}+\beta_{3}E_{3})^{\alpha}$
in many runs. This is a much less demanding task compared to a three-qubit
tomography that involves many non local measurements.

These results show that different values of $\alpha$ contain different
physical information. Moreover, it raises the interesting question
of the existence of sufficient conditions for guaranteed detection
of any heat leak.

\subsection{Bounds on system-environment correlation in a dephasing process\label{subsec: dephasing}}

As an additional example for the usefulness of the additional inequalities
(\ref{eq: alpha GPI}), let us consider the case where a system is
coupled to the environment via a dephasing interaction. A dephasing
interaction $H_{I}$ satisfies $[H_{I},H_{sys}]=0$ and as a result
it cannot affect the energy populations of the system. We also assume
that the interaction does not change the energy populations of the
environment $[H_{I},H_{env}]=0$. In summary, the reduced energy distributions
of both the system and environment are conserved (i.e., all energy
moments are conserved). 

\begin{figure}
\includegraphics[width=8.6cm]{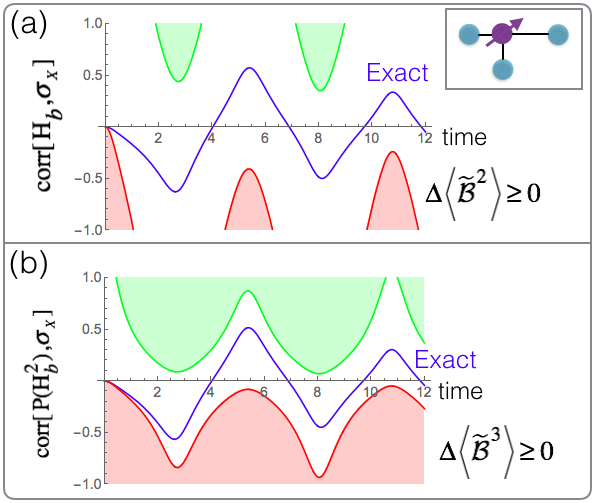}

\caption{\label{fig: corr deph}Bounds on the system-environment correlation
dynamics derived from the global passivity inequalities (\ref{eq: alpha GPI}).
(a) The blue curve depicts $\text{corr}[H_{env},\sigma_{x}]$ between
the energy of a three-spin microbath and the coherence $\left\langle \sigma_{x}\right\rangle $
of a one-spin system (see upper right box) as a function of time under
a dephasing type interaction. The red and green shaded area show lower
and upper bound on this correlation as obtained from the $\alpha=2$
global passivity inequality. (b) The $\alpha=3$ passivity inequality
leads to even tighter bound (red) on the correlation to higher energy
moments of the environment $P(H_{env}^{2})=\beta_{x}^{2}\beta\tilde{H}_{env}+(\beta H_{env})^{2}$
where $\beta$ is the temperature of the environment and $\beta_{x}$
determines the magnitude of the initial coherence of the system. }
\end{figure}

In our example, the system is a single spin with a Hamiltonian $H_{s}=\sigma_{z}$.
This spin is prepared with some initial coherence in the $x$ direction
so that $\rho_{0}^{sys}=e^{-\beta_{x}\sigma_{x}}/Z_{sys}$. As an
environment we take a three-spin microbath at inverse temperature
$\beta$ (see inset in Fig. \ref{fig: corr deph}a) and a Hamiltonian
$H_{env}=\sum_{i=1}^{3}\sigma_{z}^{(i)}$. The system and the microbath
are initially uncorrelated. The total initial density matrix is 
\begin{equation}
\rho_{0}^{tot}=e^{-\beta H_{env}\otimes I_{s}-\beta_{x}I_{env}\otimes\sigma_{x}}/Z,
\end{equation}
where $I_{env},I_{s}$ are identity operators in the Hilbert space
of the microbath and the system, and $Z$ normalizes the density matrix. 

For convenience we define a shifted version of $\hat{B}$
\begin{align}
\tilde{\mc B} & =\beta\tilde{H}_{env}+\beta_{x}\tilde{\sigma}_{x},\label{eq: Shift B}\\
\tilde{H}_{env} & =H_{env}\otimes I_{s}-E_{env,0}I,\label{eq: shift H}\\
\tilde{\sigma}_{x} & =I_{env}\otimes\sigma_{x}+1/2I,\label{eq: Shift sx}
\end{align}
where $I=I_{env}\otimes I_{s}$ is the identity operator in the joint
system-microbath space, and $E_{env,0}$ is the ground state energy
of the environment. Now $\tilde{H}_{env},\tilde{\sigma}_{env}$ and
$\tilde{\mc B}$ are all positive operators, and in addition it holds
that $\tilde{\sigma}_{x}^{n}=\tilde{\sigma}_{x}$ which will be useful
later on. This shift is needed to ensure that powers of $\tilde{\mc B}$
will remain passive w.r.t. to $\rho_{0}^{tot}$. For example, if the
operator is shifted so that it contains both positive and negative
eigenvalues, than squaring it will change the eigenvalues order and
destroy passivity w.r.t. $\rho_{0}^{tot}$. Alternatively stated $g=x^{2}$
is increasing only for $x>0$ so $\tilde{\mc B}$ must be non-negative
for even orders of (\ref{eq: alpha GPI}) to be passive.

Next we wish to find a bound on the covariance between the energy
of the microbath and the $\sigma_{x}$ values of the system at time
t, i.e., $\text{cov}[H_{env},\sigma_{x}]_{t}=\left\langle H_{env}\sigma_{x}\right\rangle _{t}-\left\langle H_{env}\right\rangle \left\langle \sigma_{x}\right\rangle _{t}$.
Since $\left\langle H_{env}\right\rangle =const$ and $\left\langle \sigma_{x}\right\rangle $
is easy to measure, the non trivial term is $\left\langle H_{env}\sigma_{x}\right\rangle _{t}$.
This term can be bounded using (\ref{eq: alpha GPI}). After some
algebra the $\alpha=2$ inequality $\Delta\left\langle (\beta\tilde{H}_{env}+\beta_{s}\tilde{\sigma}_{x})^{2}\right\rangle \ge0$
leads to
\begin{equation}
2\beta\beta_{x}\Delta\left\langle \tilde{H}_{env}\tilde{\sigma}_{x}\right\rangle \ge\beta_{x}^{2}\Delta\left\langle \tilde{\sigma}_{x}^{2}\right\rangle =\beta_{x}^{2}\Delta\left\langle \tilde{\sigma}_{x}\right\rangle .
\end{equation}
Subtracting $\left\langle H_{env}\right\rangle \left\langle \sigma_{x}\right\rangle _{t}-\left\langle H_{env}\right\rangle \left\langle \sigma_{x}\right\rangle _{0}$
we get
\begin{equation}
\text{cov}[H_{env},\sigma_{x}]_{t}\ge(\frac{\beta_{x}}{\beta}-\left\langle H_{env}\right\rangle )\Delta\left\langle \tilde{\sigma}_{x}\right\rangle .\label{eq: cov h1 sx}
\end{equation}
We now turn to consider Eq. (\ref{eq: alpha GPI}) with $\alpha=3$.
Some straightforward algebraic manipulations result in
\begin{align}
\text{cov}[P(\tilde{H}_{env}^{2}),\tilde{\sigma}_{x}]_{t} & \ge c_{0}\Delta\left\langle \tilde{\sigma}_{x}\right\rangle ,\label{eq: cov H2 sx}
\end{align}
where
\begin{equation}
P(\tilde{H}_{env}^{2})=(\beta\tilde{H}_{env})^{2}+\beta_{x}^{2}(\beta\tilde{H}_{env}),\label{eq: P Hb2}
\end{equation}
and $c_{0}=[\frac{1}{3}\beta_{x}^{2}-\left\langle \beta^{2}\tilde{H}_{env}^{2}+\beta_{x}^{2}\beta\tilde{H}_{env}\right\rangle ]$.
Due to the dephasing interaction $c_{0}$ is fixed in time and process-independent. 

Equations (\ref{eq: cov h1 sx}) and (\ref{eq: cov H2 sx}) bound
the buildup of system-environment covariance. To test these thermodynamic
bounds numerically we choose the interaction Hamiltonian $H_{I}=\sum_{i=1}^{3}\xi_{i}\sigma_{z}^{(i)}\otimes\sigma_{z}^{sys}$
where $\xi_{1}=0.7,\xi_{2}=0.5,\xi_{3}=0.3$ that mimics a coupling
that depends on the proximity of each microbath spin to the system.
The blue curve in Fig \ref{fig: corr deph}a shows the exact evolution
of the correlation function $\text{corr[\ensuremath{H_{env}},\ensuremath{\sigma_{x}}]=cov}[H_{env},\sigma_{x}]_{t}/\sqrt{Var(H_{env})Var(\sigma_{x})}$.
The red-shaded areas show the lower bound (\ref{eq: cov h1 sx}) obtained
from the $\alpha=2$ global passivity inequality. Figure \ref{fig: corr deph}b
plots the correlation function $\text{corr}[P(\tilde{H}_{env}^{2}),\sigma_{x}]=\text{cov}[P(\tilde{H}_{env}^{2}),\sigma_{x}]_{t}/\sqrt{Var[P(\tilde{H}_{env}^{2})]Var(\sigma_{x})}$
(blue curve) and the lower bound (\ref{eq: cov H2 sx}) obtained from
the $\alpha=3$ global passivity inequality.

Using the $\alpha$ global passivity inequalities (\ref{eq: alpha GPI})
and (\ref{eq: shifted reference}), it is also possible to get other
types of bounds. In the example above, by applying a $\pi$ pulse
(a unitary transformation) we obtain $\rho_{1}$ which is identical
to $\rho_{0}$ but the system spin point in the opposite direction
$\sigma_{x}\to-\sigma_{x}$. As a result, we get an \textit{upper}
bound on the system-environment covariance as shown by the green-shaded
areas in Fig. \ref{fig: corr deph}. We note that for small setups,
upper bounds can also be obtained from the hierarchy relation described
in Appendix IV.

\subsection{Detecting evasive feedback demons\label{sec: Detecting-feedback-demons}}

A feedback is an operation applied to the system of interest, where
the type of operation to be applied depends on the present state of
the system. That is, the system is first measured and then different
operations are executed depending on the measurement result. It is
well known that feedback can completely change the thermodynamics
of a process, and has to be properly accounted for. The most well-known
example is Maxwell's celebrated demon \cite{leff2014MaxwellDemonBook}.
In this thought experiment feedback is used to separate cold molecules
from hot molecules, and thereby ``violate'' the second law of thermodynamics.
A more modern viewpoint suggest the second law holds if the information
gained by the measurement is taken into account, or if the demon and
the system are treated jointly as one big setup. The thermodynamics
of information has become an active research field with many theoretical
advances \cite{parrondo2015thermoOfInfoRev}. Several experimental
realizations have also been reported \cite{toyabe2010MaxwellDemonExp,vidrighin2016photonicMaxwellExp,koski2014SzilardExp,berut2012LandauerExp}.

In this section, we study the operation of a ``lazy Maxwell's demon''.
As in Maxwell's original thought experiment the demon attempts to
create an ``anomalous heat flow'' where heat flows from the cold
environment the hot environment. Since the second law forbids such
a heat flow in the absence of a feedback or external work, the presence
of the demon can be detected if the CI (\ref{eq: Basic CI}) does
not hold. To challenge the limits of thermodynamics we assume that
our demon is lazy. Even when the conditions are such that the demon
should take action (i.e. shut the trap door) it may doze of and do
nothing. Specifically it will act a fraction $\chi$ of the times
it should. When $\chi$ is sufficiently small ($\chi\le\chi_{crit})$
no anomalous heat will be observed. Thus, the action of the demon
will not violate the CI, and the CI cannot be used to detect the demon.

Similarly to the discussion in Sec. \ref{subsec:Heat-leaks-detection}
when the CI holds we cannot exclude the presence of feedback. In particular,
there are two cases where this strategy can fail: (1) the feedback
is very weak with respect to the flows generated by the thermodynamic
protocol. (2) The elements (e.g. microbath) are initially correlated
so the CI fails even without feedback. In this latter case, we can
check for violation of the CCI (\ref{eq: ECI corr}). Yet, in this
case as well, the feedback may be too weak to violate the CCI. Therefore,
we focus here on the weak feedback scenario (small $\chi$). 

The question we pose is: can thermodynamics be used to detect lazy
demons that the CI cannot detect ($\chi<\chi_{crit}$)? Below we present
an example that demonstrates that the global passivity inequalities
(\ref{eq: alpha GPI}) can do the job. 

In quantum mechanics, feedback can be described in the following way.
The demon measures an observable with outcome $k$. For simplicity,
we consider standard projective measurements (see \cite{AmikamFlywheel}
for feedback with weak measurement in thermodynamics). The measurement
operation is described by a projection operator $\Pi_{k}^{2}=\Pi_{k}$.
Depending on the result $k$ the experimentalist applies different
unitaries $V_{k}$ . The final density matrix with feedback is given
by
\begin{equation}
\rho_{f}^{tot}=\sum_{k'}V_{k'}\Pi_{k'}[\sum_{k}p_{k}U_{k}\rho_{0}^{tot}U_{k}^{\dagger}]\Pi_{k'}V_{k'}^{\dagger}\label{eq: basic feedback-1}
\end{equation}
where the expression in the square bracket is $\rho_{f}^{tot}$ just
before the feedback. While both Eq. (\ref{eq: rho f mixU}) and Eq.
(\ref{eq: basic feedback-1}) are Kraus maps, Eq. (\ref{eq: basic feedback-1})
is \textit{not} a mixture of unitaries due to the presence of the
measurement projectors. Hence, the inequalities (\ref{eq: alpha GPI})
are guaranteed to hold for (\ref{eq: rho f mixU}) while for (\ref{eq: basic feedback-1})
they may be violated ($\Delta\left\langle \mc B^{\alpha}\right\rangle <0$).
This can be used to detect whether a feedback has been applied to
the system. 

Obviously, if for a specific $\alpha$ the inequality $\Delta\left\langle \mc B^{\alpha}\right\rangle \ge0$
holds, the presence of a demon cannot be excluded, and other $\alpha$
values should be checked. The interesting question about the existence
of a sufficient set of tests to detect any feedback is outside the
scope of the present paper and will be explored elsewhere. Our more
modest goal here is simply to demonstrate that in some cases the global
passivity inequalities (\ref{eq: alpha GPI}) are more sensitive compared
to the CI, and can detect weak feedback that the CI cannot detect.

\begin{figure}
\includegraphics[width=8.6cm]{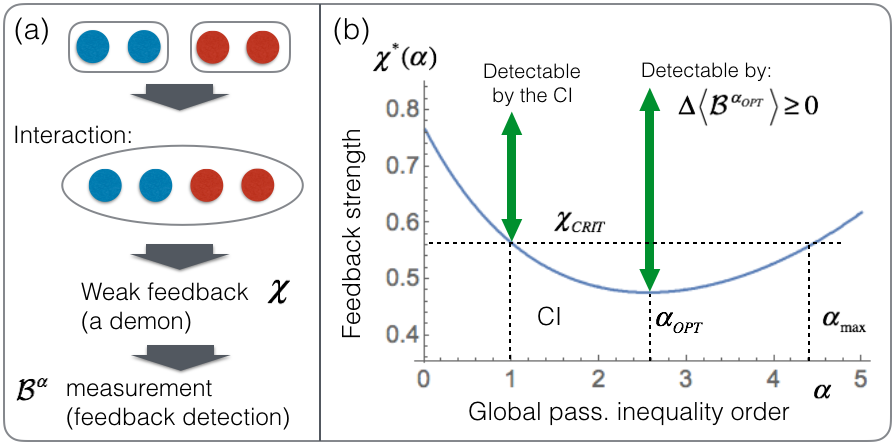}

\caption{\label{fig: feedback}(a) A schematic description of a process with
a weak feedback (``lazy demon''). A hot and cold microbaths are
coupled, and evolve unitarily for a finite time. Then a feedback (see
text) is applied to the setup with probability $\chi$ (feedback strength).
(b) Lazy demon detection using global passivity inequalities (\ref{eq: alpha GPI}).
When the feedback strength exceeds $\chi_{crit}$ the heat changes
its sign, and flows from the (initially) cold qubits to the hot qubits.
In this case the action of the demon can be detected by a violation
of CI. For $\chi\le\chi_{crit}$ the CI cannot detect the demon. The
blue curve depicts the value of $\chi^{*}(\alpha)$ for which $\Delta\left\langle B^{\alpha}\right\rangle =0$.
The global passivity inequalities (\ref{eq: alpha GPI}) can detect
the demon for $\chi\ge\chi^{*}(\alpha)$ (right-green arrow). Since
for $1\le\alpha\le\alpha_{max}$, $\chi^{*}(\alpha)\le\chi_{crit}$
we conclude that global passivity can be used to detect demon that
CI cannot. Interestingly, the best detection takes place at a non
integer value $\alpha_{opt}\simeq2.6$. }
\end{figure}

To illustrate how the global passivity inequalities (\ref{eq: alpha GPI})
can assist in feedback detection, we study a simple setup of two microbaths
composed of two spins each (Fig. \ref{fig: feedback}a). The cold
(hot) microbath is initially prepared in temperature $T_{c}=1.5$
($T_{h}=2.5$). The Hamiltonian of the setup before the interaction
is $H_{0}=\sum_{i=1}^{4}\sigma_{z}^{(i)}$. The initial state of the
setup is $\rho_{0}^{tot}=\exp[-\beta_{c}(\sigma_{z}^{(1)}+\sigma_{z}^{(2)})-\beta_{h}(\sigma_{z}^{(3)}+\sigma_{z}^{(3)})]/Z.$
At $t=0$ an 'all to all' interaction is turned on: $H_{I}=\sum_{i>j}\sigma_{+}^{(i)}\sigma_{-}^{(j)}+\sigma_{-}^{(i)}\sigma_{+}^{(j)}$
. Such ``all to all'' coupling can be realized in ion traps or in
superconducting qubits. The system evolves until $t=1$ under the
Hamiltonian $H=H_{0}+H_{I}$. After the evolution, the demon is awake
with probability $\chi$ and then it measures the setup. If the demon
finds the setup to be in the state $\ket{\uparrow_{c}\uparrow_{c}\downarrow_{h}\downarrow_{h}}$
it changes it to $\ket{\downarrow_{c}\downarrow_{c}\uparrow_{h}\uparrow_{h}}$
(where $\uparrow$ and $\downarrow$ stand for spin up or spin down
states). In all other cases, the demon does nothing. This operation
is energy conserving and transfers energy (heat) from the cold microbath
to the hot microbath. Thus, no work is involved in applying this feedback.
The process is shown in Fig. \ref{fig: feedback}a. Numerically, we
apply the feedback and check if it is detectable. For $\chi=0$ (the
demon never does anything) heat flows to the cold microbath, and the
dynamics is consistent with the CI. Thus for small values of $\chi$
this very subtle form of feedback may not be detectable by the CI. 

Since our setup includes two microbaths and no system, the inequality
(\ref{eq: alpha GPI}) with $\alpha=1$ is identical to the CI (\ref{eq: CI no sys}),
$\beta_{c}q_{q}+\beta_{h}q_{h}\ge0$. We find that for $\chi$ values
exceeding $\chi_{crit}\simeq0.56$ heat flows from the cold microbath
to the hot microbath and the demon is detectable by the CI (\ref{eq: CI no sys}).
For other $\alpha$ values we denote by $\chi^{*}(\alpha)\doteq\min_{\chi}(\Delta\left\langle \mc B^{\alpha}\right\rangle <0)$
the smallest feedback strength that leads to violation of $\Delta\left\langle \mc B^{\alpha}\right\rangle \ge0$.
The blue curve in Fig. \ref{fig: feedback}b, shows $\chi^{*}(\alpha)$.
Interestingly, this curves is not necessarily monotonically decreasing
with $\alpha$. For this feedback operation, we find that among the
$\alpha$ inequalities (\ref{eq: alpha GPI}) there is an optimal
value $\alpha_{OPT}\simeq2.6$ where the largest detection range is
observed (right green arrow). In this example, the best detection
takes place for a non integer value of $\alpha$. This gives a justification
for studying also fractional values of $\alpha$. 

Figure \ref{fig: feedback}b shows that the global passivity inequalities
(\ref{eq: alpha GPI}) successfully detect ``lazy demons'' that
the CI cannot detect. As in Sec. \ref{subsec: basic PSL1} these results
demonstrate that different values of $\alpha$ contain different physical
information. 

\section{Conclusion}

The second law of thermodynamics is one the pillars of theoretical
physics with countless applications in engineering and science. It
was originally formulated for macroscopic systems and reservoirs where
one could assume weak coupling, extensivity, and lack of recurrences.
Under some restrictions, the second law can also be applied to microscopic
systems and microscopic environments. However, these restrictions
exclude important microscopic setups. In other microscopic scenarios
(e.g. dephasing), the second law is valid but provides trivial information
that is of little use.

In this paper, we have introduced the notion of global passivity.
We have shown that under the standard thermodynamic assumptions global
passivity (complemented by passivity-divergence relation) recovers
the Clausius inequality formulation of the second law. We then show
that global passivity leads to a modified Clausius inequality that
remains valid even when the system and environment are initially strongly
correlated. This extension is important since for small systems the
interaction between system and environment and the resulting initial
correlations are in general non-negligible and the standard second
law (in particular the Clausius inequality form) cannot be used.

Our second main finding is a continuous family of global passivity
inequalities that accompany the second law. We have demonstrated how
they can be used to detect heat leaks and feedback (demons) in cases
where the second law fails to do so. In addition, the same inequalities
were used to put upper and lower bounds on the buildup of system-environment
covariance in a dephasing scenario. Such predictions are presently
outside the scope of other modern thermodynamic frameworks (e.g.,
stochastic thermodynamics, and thermodynamic resource theory).

Beyond the two main findings of this paper, the global passivity formalism
presents a set of tools that impose restrictions on observables in
thermodynamic processes in quantum systems. As an example, in a different
publication, we will study the intimate relation between passivity-divergence
relation and quantum coherence measures. Further research is needed
in order to identify which inequalities are the most useful for a
given setup. 

The present predictions of global passivity are relevant to present
day experimental setups such as ion traps, superconducting circuits,
optical lattices, and NMR. 
\begin{acknowledgments}
This work was supported by the U.S.-Israel Binational Science Foundation
(Grant No. 2014405), by the Israel Science Foundation (Grant No. 1526/15),
and by the Henri Gutwirth Fund for the Promotion of Research at the
Technion. We thank Prof. C. Jarzynski for pointing out the relation
between the CCI in the couple thermal state, and a recent classical
results for a similar scenario \cite{Jarzynski2017_PRX_strong_coupling}.
\end{acknowledgments}

\section*{Appendix I - the all observable analog of the Clausius inequality\label{subsec: basic PSL1}}

Let us start with the standard initial condition of thermodynamic
setups (assumption 1\&2 in Sec. \ref{sec: The CI}) 
\begin{equation}
\rho_{0}^{tot}=\rho_{0}^{sys}\otimes\rho^{-\sum\beta_{k}H_{k}}/Z,\label{eq: std IC}
\end{equation}
where $Z$ is a normalization factor, $H_{k}$ is the Hamiltonian
of microbath $k$, and $\beta_{k}$ is its inverse temperature at
time zero. Substituting (\ref{eq: std IC}) in the global passivity
inequality (\ref{eq: basic PI}) gives
\begin{align}
\Delta\left\langle \mc B^{sys}\right\rangle +\sum_{k}\beta_{k}q_{k} & \ge0,\label{eq: PSL1}
\end{align}
where
\begin{equation}
\mc B^{sys}\doteq-\ln\rho_{0}^{sys}.
\end{equation}
Equation (\ref{eq: PSL1}) is the observable-only analog of the CI.
It contains only linear terms in the final density matrix. The CI
and Eq. (\ref{eq: PSL1}) both reduce to $\sum_{k}\beta_{k}q_{k}\ge0$
in two important cases: 1) when there is no mediating system (e.g.
in absorption refrigerators with a tricycle interaction \cite{hofer2016SCtricycle,Roulet_3_Ion_fridge2017}).
2) In (quasi) stationary periodic operation of a heat machine interacting
with a large bath, where $\Delta\left\langle \mc B^{sys}\right\rangle $
is negligible with respect to the accumulated heat exchanges. 

Despite the similarities, Eq. (\ref{eq: PSL1}) and the CI differ
on a fundamental level. The Eq. (\ref{eq: PSL1}) contains only changes
in expectation values of operators. Alternatively stated, it is linear
in the final density matrix $\rho_{f}^{tot}$. It is nonlinear in
$\rho_{0}^{tot}$ but this is due to the fact that the observable
$\mc B$ is constructed from $\rho_{0}^{tot}$. In contrast, the term
$\Delta S^{sys}$ in the CI does not have an expectation value structure
and it is a non-linear function of $\rho_{f}^{tot}$. 

Even though the CI and Eq. (\ref{eq: PSL1}) involve different quantities,
they can be quantitatively compared, as both provide a prediction
on the quantity $\sum\beta_{k}q_{k}$. Using the identity (\ref{eq: rel ent ident})
for the system density matrix, Eq. (\ref{eq: PSL1}) can be recast
as
\begin{equation}
\Delta S^{sys}+\sum_{k}\beta_{k}q_{k}\ge-D(\rho_{f}^{sys}|\rho_{0}^{sys}).\label{eq: info PSL1}
\end{equation}
Since the r.h.s. of is negative it becomes evident that the CI (\ref{eq: Basic CI})
is tighter inequality compared to Eq. (\ref{eq: PSL1}). On the other
hand, it is important to emphasize that Eq. (\ref{eq: PSL1}) takes
less resources to evaluate in comparison to the CI. To evaluate the
$\Delta S^{sys}$ term in the CI a full system tomography is needed.
In contrast, in Eq. (\ref{eq: PSL1}) we measure one observable of
the system $\hat{B}^{sys}$. Only the diagonal elements of the density
matrix in the basis of $\hat{B}^{sys}$ are needed. Moreover, we do
not need to know explicitly the diagonal elements of $\rho^{sys}$;
$\left\langle \mc B^{sys}\right\rangle $ can be directly measured
without evaluating the probability distribution in the $\mc B^{sys}$
basis. Thus, in terms of utility for experiments, Eq. (\ref{eq: PSL1})
is easier to check compared to the CI. That said, the CI offers some
insight on the relationship between information (entropy) and energy
that is absent in Eq. (\ref{eq: PSL1}). 

\section*{Appendix II - Observables inspired by passivity}

The CCI (\ref{eq: CCI}) is given in terms of expectation values that
are determined by the initial density matrix. When the initial density
matrix is related to some observable, the CCI expectation values will
also be related to this observable. For example, in thermodynamics,
the microbaths are determined by their initial temperature and their
Hamiltonian so $\mc B^{bath}$ or $\mc B^{tot}-\mc B^{sys}$ can be
related to the Hamiltonian. In this appendix, we address the general
case where there is no prior knowledge about the operator associated
with the setup preparation. We limit our discussion to two initially
correlated parties. The extension to more parties is trivial. 

$\mc B^{tot}=-\ln\rho_{0}^{tot}$ is a positive Hermitian operator,
and it can be decomposed in the following way:

\begin{align}
\mc B^{tot} & =[\sum_{i=1}^{L_{A}}r_{A,i}Z_{i}\otimes\frac{I_{N_{B}\times N_{B}}}{N_{B}}+\nonumber \\
 & \sum_{j=1}^{L_{A}}r_{B,i}\frac{I_{N_{A}\times N_{A}}}{N_{A}}\otimes Z_{j}+\sum_{i=1}^{L_{A}-1}\sum_{j=1}^{L_{B}-1}t_{ij}Z_{i}\otimes Z_{j}],\label{eq: rho init}
\end{align}
where $L_{A(B)}=N_{A(B)}^{2}-N_{A(B)}$, and $\{Z_{i(j)}\}_{1}^{L_{A(B)}-1}$
are \textit{traceless} orthonormal basis operators for $N_{A(B)}\times N_{A(B)}$
Hermitian traceless matrices. $Z_{\ensuremath{L_{A(B)}}}=I_{N_{A(B)}\times N_{A(B)}}$
are the identity operators in each party. Finally $r_{A}$ and $r_{B}$
determine the reduced density matrices: 
\begin{equation}
\text{tr}_{B}\mc B^{tot}=[\sum r_{A,i}Z_{i}]=\mc B_{A}.
\end{equation}

This suggest that \textbf{$\mc B^{tot}=\mc B_{A}+\mc B_{B}+\mc B_{int}$
}where $\mc B_{int}$ has an interaction Hamiltonian form $\mc B_{int}=\sum_{ij}t_{ij}Z_{i}\otimes Z_{j}$.
Now let A be the system and B be the environment. We find that CCI
has the form:
\begin{equation}
\Delta S^{sys}+\Delta\left\langle \mc B^{bath}\right\rangle +\Delta\left\langle \mc B^{int}\right\rangle +\Delta\left\langle \mc B^{sys}-\mc B_{\text{eff}}^{sys}\right\rangle \ge0
\end{equation}
where
\begin{align}
\mc B^{bath} & =tr_{sys}(-\ln\rho^{tot}),\\
\mc B^{sys} & =tr_{env}(-\ln\rho^{tot}),\\
\mc B_{\text{eff}}^{sys} & =-\ln[tr_{env}(\rho^{tot})].
\end{align}
Thus the dressing term originate from the non commutativity of the
\textit{partial} trace operation and the $-\ln$ operation. The two
commute when $\rho_{0}^{tot}$ is in a system-environment product
state.

\section*{Appendix III - the CCI for two simple interactions in a coupled thermal
state}

In this appendix, we evaluate the magnitude of the system dressing
term in the CCI with an initially coupled thermal state (see Sec.
\ref{subsec:A-joint-system-bath}) $\beta(H_{s}-H_{\text{eff}}^{sys})$
for two interesting cases. First, we make the simplifying assumption
that $H_{I,0}$ conserves the sum of bare energy of the system and
the microbath
\begin{equation}
[H_{I,0},H_{s}+H_{\mu b}]=0.
\end{equation}
Using this condition we find

\begin{align}
\text{tr}{}_{h}e^{-\beta(H_{s}+H_{I,0}+H_{b})} & =e^{-\beta H_{s}}\text{tr}{}_{h}[e^{-\beta H_{b}}e^{-\beta H_{I,0}}]\nonumber \\
 & =e^{-\beta H_{s}}\sum_{E_{b}}e^{-\beta E_{b}}\braOket{E_{b}}{e^{-\beta H_{I,0}}}{E_{b}}\nonumber \\
\end{align}
where in the last stage we have used the microbath energy eigenstates
to do the partial trace. 

We consider two typical forms of $H_{I,0}$: (1) $\epsilon\mc H_{s}\otimes\mc H_{b}$
where $[\mc H_{b},H_{b}]=[\mc H_{s},H_{s}]=0$ which often arises
in dephasing environments, e.g. $\sigma_{z}\otimes\sigma_{z}$. (2)
a swap-like interaction between the system and the microbath $\epsilon(a_{b}^{\dagger})\otimes a_{s}+h.c.$
where $a$ is some transition operator of the form $\ketbra mn$. 

For case (1), simple dephasing, we find
\begin{align}
\text{tr}{}_{h}e^{-\beta(H_{s}+H_{h}+\epsilon\mc H_{s}\otimes\mc H_{b})} & =e^{-\beta H_{s}}\nonumber \\
 & \sum_{n=0}\frac{(-\beta\epsilon\mc H_{s})^{n}\otimes tr[e^{-\beta H_{b}}\mc H_{b}^{n}]}{n!}\nonumber \\
= & e^{-\beta H_{s}}\sum_{n=0}\frac{(-\beta\epsilon\mc H_{s})^{n}}{n!}\left\langle \mc H_{b}^{n}\right\rangle _{0}Z_{b0},\label{eq: rho_sys_deph}
\end{align}
where $Z_{b0}=\text{tr}[e^{-\beta H_{b}}]$ and $\left\langle \mc H_{b}^{n}\right\rangle _{0}=\text{tr}[\frac{e^{-\beta H_{b}}}{Z_{b0}}\mc H_{b}^{n}]$.
We define the function $f(x)=\sum_{n=0}\frac{Z_{b0}\left\langle \mc H_{b}^{n}\right\rangle _{0}}{n!}x^{n}$
so from (\ref{eq: ECI pre_fixed coupling}) and (\ref{eq: rho_sys_deph})
we get

\begin{align}
\beta\Delta(H_{\text{eff}}^{sys}-H_{s}) & =\ln f(-\beta\epsilon\mc H_{s})\nonumber \\
 & =(\beta\epsilon)\mc H_{s}\left\langle \mc H_{b}\right\rangle _{0}Z_{b0}+O[(\beta\epsilon)^{2}]\label{eq: dephasing energy term}
\end{align}
Since $\mc H_{s}$ commute with $H_{s}$ we conclude that $\ln f(-\beta\epsilon\mc H_{s})$
is a ``dressing effect'' of the microbath on the system (this term
has the same eigenstates as $H_{s}$). In the CCI (\ref{eq: CCI}),
the system appears \textit{primarily} in $\Delta S^{sys}$ as in the
CI. However, in the CCI the dephasing dressing effect contributed
a new expectation value term (in contrast to information term) of
the \textit{system} (\ref{eq: dephasing energy term}). 

For case (2), the creation-annihilation swap-like interaction we find,
\begin{align}
\text{tr}{}_{b}e^{-\beta(H_{s}+H_{I,0}+H_{b})} & =e^{-\beta H_{s}}\nonumber \\
 & \sum_{E_{b}}e^{-\beta E_{b}}\braOket{E_{b}}{e^{-\beta H_{I,0}}}{E_{b}}\nonumber \\
 & =e^{-\beta H_{s}}\nonumber \\
 & \sum_{E_{b}}e^{-\beta E_{b}}\braOket{E_{b}}{\cosh(\beta H_{I,0})}{E_{b}}.
\end{align}
The $\text{\ensuremath{\cosh}}$ appears since all the odd powers
of $e^{-\beta H_{I,0}}$ are eliminated by the partial trace. From
this we conclude that contribution of the swap term is $O[(\beta\epsilon)^{2}]$.
Therefore, if there is a weak swap term at time zero, it will manifest
in the first order in the $\beta\Delta\left\langle H_{I,0}\right\rangle $
term of the CCI, and only in second order in the system dressing term
of the CCI.

\section*{Appendix IV - $\protect\mc B^{\alpha}$ hierarchy relations}

In this Appendix we point out that there is a hierarchy relation between
inequalities (\ref{eq: alpha GPI}) with different values of $\alpha$.
The derivation is based on the fact that for any $q>p>0$ the function
$h(x)=\frac{x^{p}}{p}-\frac{x^{q}}{q}$ is monotonically increasing
for $x\in[0,1]$. To apply this to the operator $\mc B=-\ln\rho_{0}^{tot}$
we define the operator $\tilde{\mc B}=\mc B/\nrm{\mc B}_{op}$ where
$\nrm{\mc B}_{op}$ is the operator norm, which is equal to the largest
eigenvalue of $\mc B$ ($\mc B$ is Hermitian and bounded). Based
on this construction $h(\tilde{\mc B})$ is passive w.r.t to $\rho_{0}^{tot}$
since we ensured that the spectrum of $\tilde{\mc B}$ is in $[0,1]$
where $h(x)$ is an increasing function. Thus from passivity we get
that $\Delta h(\tilde{\mc B})\ge0$ which means that
\begin{align}
\frac{\Delta\left\langle \mc B^{p}\right\rangle }{p\nrm{\mc B}_{op}^{p}} & \ge\frac{\Delta\left\langle \mc B^{q}\right\rangle }{q\nrm{\mc B}_{op}^{q}},\:\:\:\forall p>q>0.\label{eq: hierarchy}
\end{align}
In particular for integer $p$ and $q$ we get
\begin{equation}
\frac{\Delta\left\langle \mc B\right\rangle }{\nrm{\mc B}_{op}}\ge\frac{1}{2}\frac{\Delta\left\langle \mc B^{2}\right\rangle }{\nrm{\mc B}_{op}^{2}}\ge\frac{1}{3}\frac{\Delta\left\langle \mc B^{3}\right\rangle }{\nrm{\mc B}_{op}^{3}}\ge..\ge\frac{1}{n}\frac{\Delta\left\langle \mc B^{n}\right\rangle }{\nrm{\mc B}{}_{op}^{n}}\ge0.
\end{equation}
These relations can be useful for bounded setups (e.g. collection
of spin) as long as the operator $\mc B$ is bounded and not too large
(e.g. when the setup include a small number of spins). For large $\nrm{\mc B}_{op}$
the hierarchy relations (\ref{eq: hierarchy}) reduce to (\ref{eq: alpha GPI}).

\bibliographystyle{apsrev4-1}
\bibliography{/Users/raam_uzdin/Dropbox/RaamCite}

\end{document}